\documentclass[prb,10pt]{revtex4}

\usepackage{amsmath}   
\usepackage{graphicx}   
\usepackage{subfigure}  
\usepackage{hyperref}  
\usepackage{amsmath}
\usepackage{amssymb}

\newcommand{\dd}{\mathrm{d}}
\renewcommand{\figurename}{Fig.\ }

\begin{document}

\title{Large Deviations of Correlation Functions in Random Magnets}
\author{F. Morone}
\affiliation{Dipartimento di Fisica, Universit\`a ``La Sapienza'', P.le A. Moro 2, I-00185, Rome, Italy}
\author{G. Parisi}
\affiliation{Dipartimento di Fisica, INFN -- Sezione di Roma1, CNR-IPCF UOS Roma Kerberos, Universit\`a ``La Sapienza'', P.le A. Moro 2, I-00185, Rome, Italy}
\author{F. Ricci-Tersenghi}
\affiliation{Dipartimento di Fisica, INFN -- Sezione di Roma1, CNR-IPCF UOS Roma Kerberos, Universit\`a ``La Sapienza'', P.le A. Moro 2, I-00185, Rome, Italy}
\begin{abstract}
We present a large deviations theory of the spin-spin correlation functions in the Random Field Ising Model on the Bethe lattice, both at finite and zero temperature. Rare events of atypically correlated variables are particularly important at the critical point: the phase transition is driven by few pairs of strongly correlated spins, while the majority remains basically uncorrelated. At the zero temperature critical point the number of spin pairs correlated over a distance $\ell$ is shown to be no longer exponential, but only linear in the spins separation. 
\end{abstract}

\maketitle

\section{Introduction}
The presence of quenched randomness in ferromagnetic systems, and generally in all disordered systems,  induces large fluctuations in the physical observables that strongly depend on the local environment\cite{Derrida,vanHemmen}.
These fluctuations may reduce the physical informations contained in the typical  values of observables, making unavoidable the knowledge of their full probability distribution. 
The theoretical need to deal with distributions, rather than averages, arises from an important physical truth about random systems, i.e., the fact that they are not just dirty or imperfect realizations of regular systems, but fundamentally different ones\cite{Phil}.   

We undertake this important and conceptually deep approach to study the spin-spin connected correlation function, whose probability distribution remains non-trivial even in the thermodynamical limit. The most probable value of the correlation does not play any significant role, since it can be very different from the more important average value. We find that, at any given distance, the average correlation is dominated by few atypical highly correlated variables. Therefore one can not just be satisfied with computing the most likely outcomes of a measure of the correlation between the dynamical variables: one also need to understand the rare events, that is atypically large correlations appearing with very small probabilities. This is especially important at the phase transition point, where the system gets critical thanks to the presence of rare and highly correlated spin pairs\cite{Takahashi2010}. 
The study of the fluctuations of the spin-spin correlation function thus leads, in a natural way, to a theory of large deviations.

Besides the fact that no theoretical description of physical phenomena can be considered satisfactory if it does not quantify deviations and fluctuations, the non trivial distribution of correlations plays a central role in disordered systems, because it encapsulates most of the physical information, that, on the contrary, is missed by their typical values.  
Furthermore the outcome of this work is particularly important in the development of a loop expansion around the mean field theory for disordered systems, since one needs to know very precisely (i.e. at the large deviations level) the two-point correlation function in order to compute loop corrections. 
To be precise we have to specify what kind of mean field theory we are talking about. Generally, mean field approximations are formulated in two different ways: the first one consists by replacing coupling constants decaying with distance by distance-independent interactions. The second one amounts to neglect the correlations induced by the presence of loops on the lattice. The latter approximation scheme is the Bethe-Peierls (BP) approximation\cite{Bethe}. While the systematic field-theoretical treatment of the corrections to the first method is well developed and widely used\cite{DeDom} (even if we do not know how to treat the loop expansion in the spin glass phase below $6$ dimensions), the loop corrections to BP results are not under the same control. The development of a perturbative formalism  around the Bethe approximation has been undertaken in the last years using different approaches\cite{Cherkov,ParisiSlanina,MontanariRizzo}, but the computations remain analytically very challenging. In particular in the systematic expansion proposed in Ref.\cite{ParisiSlanina} it is stressed that the practical calculation of the diagrams relies on the traslational symmetry of the model. The application of a systematic loop expansion to  disorderded systems requires the precise knowledge of the fluctuations of the propagators, i.e., their full probability distribution. In this work we provide a way to numerically compute the distribution of bare propagators for a finite temperature, as well as their analytical form in the zero temperature limit (that is the most interesting case from the point of view of the renormalization group approach).

To make our discussion concrete we consider the Random Field Ising Model (RFIM) on the Bethe lattice and develop the large deviations theory of the connected two-point correlation functions. The treatment of the finite temperature regime is separated from the zero temperature one. The latter case deserves particular care, since the naive limit $T\rightarrow 0$ of the finite temperature scenario, does not account for the whole spectrum of possible zero-temperature fluctuations.

In the end of the manuscript we argue that the zero temperature results for the RFIM are also valid for  the spin glass model in a uniform field, when the system is above the critical point. On the contrary, below the critical point, the analysis of the correlations distribution in the spin glass model requires a much more sophisticated analysis.

The paper is organized as follows. In Section II we introduce the model and define the two main objects to be studied: the cumulant generating function of the decay rate of the two-point connected correlation function and the large deviations function of this decay rate. In Section III and IV we present two methods to compute these functions. The first one is the same used in Ref.\onlinecite{Takahashi2010}, while the second method allows to obtain the cumulant generating function from the solution of an integral eigenvalues equation, thus providing both numerical and theoretical advantages with respect to first method. We use both methods to draw the profiles of the large deviations function at finite temperature, with a special attention to the critical line. 
Section V is devoted to the zero temperature case. In this limit situation an analytical form of the correlation probability distribution can be derived under mild approximations. A conclusion is to be found at the end of this paper.

\section{The model}
\begin{figure}
\includegraphics[width=0.95\textwidth]{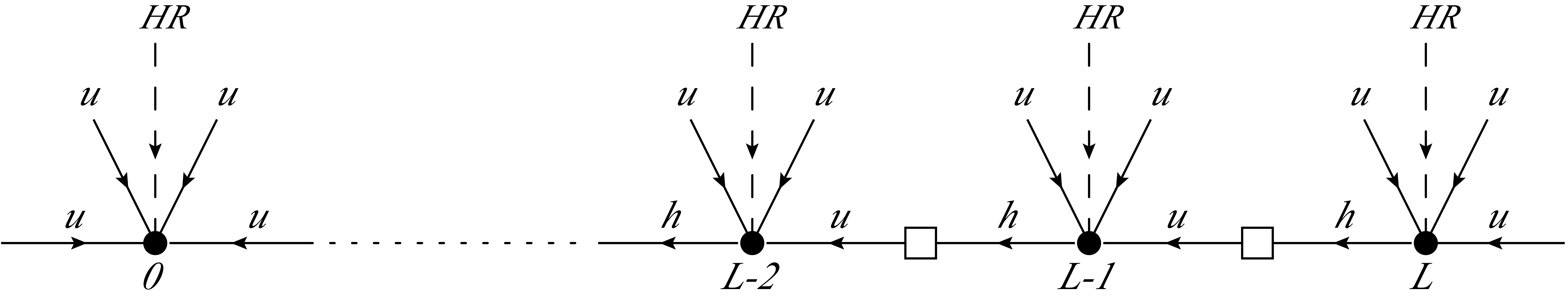}
\caption{The chain in the factor graph used to compute the response function given by Eq. \eqref{eq:Resp}
\label{fig:path}: dots represent the variable-nodes, squares the factor-nodes. In this case the factor-nodes are in one-to-one correspondence with the edges of the original graph. The cavity field $h$ is the message sent from variable-node to  factor-node, while the bias $u$ is the message from factor-node to variable-node. The external random field, called $\mathrm{HR}$ in this figure, is represented as a dashed arrow. 
 Eq. \eqref{eq:Resp} can be justified by noticing that the cavity biases $\{u\}$ entering a given node are supposed to be uncorrelated under the hypothesis of  the replica-symmetric cavity method. Therefore, deriving the field $u_{1\to0}$ with respect to the bias $u_{\to\ell}$ arriving at the right boundary of the path, and applying the chain rule for derivatives, one gets the product over $k$ of Eq. \eqref{eq:Resp}. The remaining prefactor $\beta(1-\langle s_0\rangle^2)$ comes from the derivative of the magnetization $\langle s_0\rangle$ of site $0$ with respect to a field acting on the same site $0$ (it can be the total field or just the bias $u_{1\to 0}$, both would lead to the same result).
}
\end{figure}

The Random Field Ising Model (RFIM) on the Bethe lattice is defined by the following Hamiltonian:
\begin{equation}
H = -\sum_{\langle i j\rangle}Js_is_j-\sum_i h_is_i\,\, ,
\end{equation}
where the $N$ spins $s_i = \pm1$ are placed on the nodes of a Bethe lattice with fixed coordination number $z+1$, i.e. a random regular lattice. The random fields $\{h_i\}$ are drawn independently from a Gaussian distribution with zero mean and variance $\overline{h^2}=\sigma^2$. 
The model can be exactly solved at the replica-symmetric level, since no replica-symmetry breaking does occur at any temperature~\cite{NoglassRFIM}. In the cavity approach the model solution is obtained from the following self consistent equations for the cavity field and cavity bias probability density functions $P(h)$ and $Q(u)$ respectively:
\begin{equation}
\begin{aligned}
&P(h)=\int\left[\prod_{i=1}^{z}\dd Q(u_i)\right]\overline{\delta\left(h-h_R-\sum_{i=1}^{z}u_i\right)}^{h_R}\, \label{eq:Pcav},\\ 
&Q(u)=\int \dd P(h)\ \delta\left[u-\hat{u}(\beta,J,h)\right]\, ,
\end{aligned}
\end{equation}
where the overbar denotes the average over the random field distribution and the function $\hat{u}(\beta,J,x)$ is given by the following expression:
\begin{equation}
\hat{u}(\beta,J,x) = \beta^{-1}\text{atanh}[\tanh(\beta J)\tanh(\beta x)]\ .\label{eq:uhat}
\end{equation}
The response of the spin $s_i$ with respect to a field $H_j$ acting on site $j$ is defined as
\begin{equation}
\mathcal{R}(i,j)=\frac{\partial\langle s_i\rangle}{\partial H_j}=\beta\langle s_is_j\rangle_\textrm{c},
\end{equation}  
where $\langle s_is_j\rangle_\textrm{c}$ is the connected correlation function between spins $s_i$ and $s_j$.
Let us focus on two spins $s_i$ and $s_j$ in the graph at distance $\ell$. In the thermodynamical limit the spins are joined almost surely by a unique path of lenght $\ell$. Renaming the spins at the boundaries of the path as $s_0$ and $s_\ell$, the response function can be written using cavity messages as \cite{Takahashi2010}
\begin{equation}
\mathcal{R(\ell)}=\frac{\partial\langle s_0\rangle}{\partial H_{\ell}}=\beta(1-\langle s_0\rangle^2)\prod_{k=1}^{\ell}\frac{\partial u_{k\rightarrow k-1}}{\partial u_{k+1\rightarrow k}},\label{eq:Resp}
\end{equation} 
where $u_{k\rightarrow k-1}$ is the cavity bias running along the path from site $\ell$ to site $0$ (see \figurename\ref{fig:path}).

The decay properties of the response function $\mathcal{R(\ell)}$, at large $\ell$, are contained in the product entering in the r.h.s. of Eq. \eqref{eq:Resp}. Thus we define a normalized connected correlation function as
\footnote{The word normalized is used to mean that the connected correlation function $\langle s_0 s_{\ell}\rangle_{c}$ is normalized with respect to the connected autocorrelation function $\langle s_0 s_{0}\rangle_{c}=1-\langle s_0\rangle^2$, i.e. we define $\mathcal{C}(\ell)=\langle s_0 s_{\ell}\rangle_{c}/(1-\langle s_0\rangle^2)$. }
\begin{equation}
\mathcal{C}(\ell) = \frac{\mathcal{R}(\ell)}{\beta(1-\langle s_0\rangle^2)}=\prod_{k=1}^{\ell}\frac{\partial u_{k\rightarrow k-1}}{\partial u_{k+1\rightarrow k}}\ \ .\label{eq:Cnorm}
\end{equation}
With this definition we have $\mathcal{C}(0)=1$.
The large distance behaviour of the correlation function is extracted from its decay rate:
\begin{equation}
\gamma_0=-\lim_{\ell\rightarrow\infty}\frac{\log\mathcal{C}(\ell)}{\ell}\,\, .\label{eq:typRate}
\end{equation}
Due to the FKG inequality\cite{Fortuin1971}, the correlation function $\mathcal{C}(\ell)$ is positive defined, so that its logarithm is well behaved for any finite temperature. Because of the multiplicative ergodic theorem\cite{Oseledec} the limit \eqref{eq:typRate} exists with probability one. Moreover the decay rate $\gamma_0$ is a non-random variable, i.e.,
\begin{equation}
\gamma_0=-\lim_{\ell\rightarrow\infty}\frac{\overline{\log\mathcal{C}(\ell)}}{\ell}\,\, ,
\end{equation} 
where the overbar indicates the average over the ensemble $\Lambda$ of all possibile cavity bias sequences $\mu_{\ell}\equiv[u_{\ell\rightarrow\ell-1},\dots,u_{1\rightarrow0}]$. Roughly speaking, we can say that there is a subclass $\Lambda^{*}$ of typical sequences which has full measure over the ensemble of sequences $\Lambda$ and where the correlation $\mathcal{C}(\ell)$ decays exponentially with rate $\gamma_0$. For all $\mu_{\ell}\in\Lambda^{*}$ we have that: $-\lim_{\ell\rightarrow\infty}\log\mathcal{C}(\ell)/\ell=\gamma_{0}$.
Despite this, there exist very improbable sequences in $\Lambda$ which lead to a different asymptotic limit of the decay rate. They cannot change the logarithmic average \eqref{eq:typRate}, but they are relevant in the evaluation of the decay rate of the disorder averaged connected correlation function:
\begin{equation}
\gamma_1=-\lim_{\ell\rightarrow\infty}\frac{\log\overline{\mathcal{C}(\ell)}}{\ell}\,\, .
\end{equation} 
For a general discussion on this problem see, e.g., Refs.~\onlinecite{Derrida, vanHemmen, Vulpiani}.

The rate $\gamma_1$ is the quantity directly involved in the computation of the ferromagnetic susceptibility $\chi_F$. In order to compute $\chi_F$ we use the fact that in a large random sparse network, by approximating the lattice with a tree rooted on site 0, the number of spins at distance $\ell$ from the central spin $s_0$ is $(z+1)z^{\ell-1}$. This allows to write the non-trivial contribution to the ferromagnetic susceptibility $\chi_F$ as:
\begin{equation}
\chi_F=\frac{z+1}{z}\sum_{\ell=1}^{\infty}z^{\ell}\,\,\overline{\mathcal{R}(\ell)}\propto\sum_{\ell=1}^{\infty}z^{\ell}\,\,\overline{\mathcal{C}(\ell)}\sim\left(1-ze^{-\gamma_1}\right)^{-1}\,\, .
\end{equation}
The critical point, associated with the divergence of $\chi_F$, is reached when $\gamma_1=\log(z)$. 
We now introduce the distribution of the decay rate $P_{\ell}(\gamma)$. Physically $P_{\ell}(\gamma)\dd\gamma$ represents the probability to measure an effective decay rate $-\log\mathcal{C}(\ell)/\ell$ belonging to the interval $[\gamma,\gamma+\dd\gamma]$.  Since $P_{\ell}(\gamma)$ concentrates on the point $\gamma=\gamma_0$ for $\ell\rightarrow\infty$, the following asymptotic behaviour can be predicted:
\begin{equation}
P_\ell(\gamma)\approx e^{-\ell\Sigma(\gamma)}\,\,\,  \mathrm{for}\,\,\, \ell\rightarrow\infty\,\, ,\label{eq:P_L}
\end{equation}
with a rate function $\Sigma(\gamma)\geq0$. 
We will be more precise in a while, but for the moment let us evaluate the susceptibility $\chi_F$ using the asymptotic distribution \eqref{eq:P_L}:
\begin{equation}
\chi_F\propto\sum_{\ell=1}^{\infty}z^{\ell}\,\,\overline{\mathcal{C}(\ell)}\approx \sum_{\ell}z^{\ell}\int \dd\gamma\ e^{-\ell[\Sigma(\gamma)+\gamma]}\approx\sum_{\ell}z^{\ell} e^{-\ell[\Sigma(\gamma^*)+\gamma^*]}\,\, ,
\end{equation}
where in the last step the integral is evaluated via the steepest descent method and $\gamma^*$ is the solution of the saddle point equation
\begin{equation}
\frac{\partial\Sigma(\gamma)}{\partial\gamma}\Big|_{\gamma^*}=-1\,\, .\label{eq:spGamma*}
\end{equation}
The value $\gamma = \gamma^*$ selects the correlations which dominate the average $\overline{C(\ell)}$. 
If we call $\mathcal{N}_{\ell}(\gamma^*)=z^{\ell}e^{-\ell\Sigma(\gamma^*)}$ the number of spins pairs $(\sigma_0,\sigma_{\ell})$ whose correlation $C(\ell)$ decays as $e^{-\ell\gamma^*}$, we can write the susceptibility $\chi_F$ as 
\begin{equation}
\chi_F\approx\sum_{\ell}\mathcal{N}_{\ell}(\gamma^*)e^{-\ell\gamma^*}\ \ .
\label{eq:chiferro}
\end{equation}
Eq. \eqref{eq:chiferro} emphasizes that the main contribution to the susceptibility comes from rare correlations with atypically small decay rate $\gamma^*$. This is true also at the phase transition point, with the physical consequence that the criticality is driven by few pairs of strongly correlated variables, whose number is $z^{\ell}e^{-\ell\Sigma(\gamma^*)}$ and, hence, much smaller then the total number of pairs $z^\ell$.
A scaling law of the form $P_{\ell}(\gamma)\approx e^{-\ell\Sigma(\gamma)}$, where $\ell$ is assumed to be large, and $\Sigma$ is a positive function, is referred to as a large deviation principle. To make it more precise, let $x_\ell$ be a random variable indexed by the integer $\ell$, and let $P(x_\ell\in I)$ be the probability that $x_\ell$ takes on a value in a set $I$. We say that $P(x_\ell\in I)$ satisfies a large deviation principle with rate $\Sigma_I$ if the limit
\begin{equation}
\lim_{\ell\to\infty}-\frac{1}{\ell}\log P(x_\ell\in I)\ =\ \Sigma_I
\label{eq:LDlimit}
\end{equation}
exists.
Hence, if $P(x_\ell\in I)$ has a dominant exponential behavior in $\ell$, then that limit should exist with $\Sigma_I\neq0$. If the limit does not exist, then either $P(x_\ell\in I)$ is too singular to have a limit or else $P(x_\ell\in I)$ dacays with $\ell$ faster than $e^{-\ell a}$ with $a>0$. In this case, we say that $P(x_\ell\in I)$ decays super-exponentially and set $\Sigma=\infty$. The large deviation limit may also be zero for any set $I$ if $P(x_\ell\in I)$ is \textrm{sub-exponential} in $\ell$, i.e., if $P(x_\ell\in I)$ decays slower than $e^{-\ell a},\ a>0$. The cases of interest in the analysis of large deviations of correlations are those for which the limit shown in \eqref{eq:LDlimit} exists. Indeed, a sub-exponential behavior of $P(x_\ell\in I)$ would lead to an unphysical divergent susceptibility at all temperatures, while a super-exponential one would give a finite susceptibility at all temperatures, that corresponds to a model without phase transitions.

To reconstruct the rate function $\Sigma(\gamma)$, we define the cumulant generating function of the random variable $\gamma$ \cite{Touchette2009}:
\begin{equation}
\lambda(q)=-\lim_{\ell\rightarrow\infty}\frac{\log\overline{\mathcal{C}(\ell)^q}}{\ell}\,\, ,\label{eq:lambdaq}
\end{equation}
where $q\in\mathbb{R}$ and
\begin{equation}
\overline{\mathcal{C}(\ell)^q}{}=\int \dd\gamma\ P_{\ell}(\gamma)\ e^{-\ell q\gamma}\,\, .
\end{equation}
Notice that $\gamma_0=\lambda^{\prime}(0)$ and $\gamma_1=\lambda(1)$. The G{\"a}rtner-Ellis theorem\cite{Ellis,Gartner} states that, if $\lambda(q)$ exists and is differentiable for all $q\in\mathbb{R}$, then $\gamma$ satisfies a large deviation principle, i.e.,
\begin{equation}
P_{\ell}(\gamma)\approx e^{-\ell\Sigma(\gamma)}\,\, ,
\end{equation}
where the sign '$\approx$' is used to stress that, as $\ell\rightarrow\infty$, the dominant part of $P_{\ell}(\gamma)$ is the decaying exponential $e^{-\ell\Sigma(\gamma)}$. The rate function $\Sigma(\gamma)$ is given by the Legendre-Fenchel\cite{Ellis} trasform of $\lambda(q)$:
\begin{equation}
\Sigma(\gamma)=\sup_{q\in\mathbb{R}}[\lambda(q)-q\gamma]\,\, .\label{eq:ratefunction}
\end{equation}
In the next section we are going to explain how to compute the cumulant generating function \eqref{eq:lambdaq} and the rate function \eqref{eq:ratefunction}.

\section{How to compute the cumulant generating function $\lambda(q)$}
In this Section we present two methods to compute the generating function $\lambda(q)$. The first method follows from the formal definition of $\lambda(q)$. Practically we compute the average $\overline{\mathcal{C}(\ell)^q}$ at finite $\ell$ using a stochastic sampling method. Once we have obtained the average $\overline{\mathcal{C}(\ell)^q}$ we compute the logarithm, then we divide by $\ell$ and eventually  we extrapolate in the limit $\ell\rightarrow\infty$. Even if the extrapolation is unambigous from the numerical point of view, it remains limited from the theoretical perspective. It would be better to have a recipe that allows to compute directly the asymptotic function $\lambda(q)$. Fortunately this method exists and it will be described below. It will be demonstrated that the cumulant generating function $\lambda(q)$ is the solution of an integral eigenvalues equation. This equation was already present in the literature \cite{Weigt1996} and it was first derived using the replica theory. Here we give a parallel derivation, using the language of the cavity method.

\subsection{Method I}\label{subsed:metodo1}

To obtain $\lambda(q)$ we first compute the moments of the correlation function $\overline{\mathcal{C}(\ell)^q}$:
\begin{equation}
\overline{\mathcal{C}(\ell)^q}=\overline{\prod_{k=1}^{\ell}\left(\frac{\partial u_{k\rightarrow k-1}}{\partial u_{k+1\rightarrow k}}\right)^q}\,\, .\label{eq:Cformula}
\end{equation}
The variables $\{u_{k\rightarrow k-1}\}_{k=1}^{\ell}$ are correlated variables and this makes the analitycal computation hard to pursue in full generality. Neverthless the estimation of $\lambda(q)$ can be accomplished numerically with the following method. For any fixed value of $q$, one generates many instances of $\mathcal{C}(\ell)^q$ by iterating the following map:
\begin{equation}
u_{k\rightarrow k-1}=\beta^{-1}\text{atanh}\left[\tanh(\beta J)\tanh\left(\beta h_R+\beta u_{k+1\rightarrow k}+\beta\sum_{j=1}^{z-1}u_{j\rightarrow k}\right)\right],\ \ \ \text{for}\ \ k = \ell,\ell-1,\dots,1 \ ,
\end{equation}
and applying Eq. \eqref{eq:Cformula}. The variables $u_{j\rightarrow k}$ are the cavity bias coming from the $(z-1)$ branches outside the path that merge on the node $k$. Within the cavity approach they are assumed to be independent and identically distributed according to the distribution $Q(u)$ given by Eq.\eqref{eq:Pcav}. The starting cavity bias, $u_{\ell+1\rightarrow\ell}$, is also extracted from the distribution $Q(u)$.  The average $\overline{\mathcal{C}(\ell)^q}$ is obtained as the mean over the instances generated with this method. 
Defining an intermediate generating function $\lambda_\ell(q)$ as:
\begin{equation}
\lambda_\ell(q)=-\frac{\log\overline{\mathcal{C}(\ell)^q}}{\ell}\,\, ,\label{eq:lambdaintermediate}
\end{equation}
eventually we have to take the limit:
\begin{equation} 
\lim_{\ell\rightarrow\infty}\lambda_{\ell}(q)=\lambda(q)\,\, .
\end{equation}
Assuming for $\lambda_{\ell}(q)$, in the limit $\ell\rightarrow\infty$, an expansion in powers of $\ell^{-1}$ of the form: 
\begin{equation}
\lambda_{\ell}(q) = \lambda(q)+\frac{A(q)}{\ell}+o(\ell^{-1}),\label{eq:lexp}
\end{equation}
we obtain the value $\lambda(q)$ extrapolating $\lambda_{\ell}(q)$ with the function \eqref{eq:lexp}.
In practice we take $1250$ values of $q$ in the range $q\in[-0.25,6.0]$, equally spaced by $\Delta q=0.005$, and vary $\ell$ from $\ell=1$ to $\ell=10$. 
For each value of $q$ and $\ell$ we  take the average $\overline{\mathcal{C}(\ell)^q}$ over $10^7$ instances and compute $\lambda_{\ell}(q)$. We collect 100 values of the same $\lambda_{\ell}(q)$ and estimate statistical errors. In the end the extrapolation $\lambda_{\ell}(q)\rightarrow\lambda(q)$ is performed using Eq. \eqref{eq:lexp}.

\begin{table}[h]
\begin{center}
\begin{tabular}{|c|c|}
\hline
$\sigma$ & $T_c(\sigma)$\\
\hline
\ $0.00$ & $1.820478$ \\
\hline
\ $0.20$ & $1.7805(5)$ \\
\hline
\ $0.50$ & $1.5755(5)$ \\
\hline
\ $0.80$ & $1.1590(5)$ \\
\hline
\ $1.00$ & $0.5495(5)$ \\
\hline
\ $1.02$ & $0.3980(5)$ \\
\hline
\ $1.035$ & $0.1635(5)$ \\
\hline
\ $1.037(1)$ & $0.0$ \\
\hline
\end{tabular}
\caption{Some critical temperatures $T_c(\sigma)$ in the RFIM on the Bethe lattice with coordination number $z+1=3$ and Gaussian random fields. The last line corresponds to the zero temperature critical point, estimated with the method described in Section \ref{sec:ZT}.}\label{Tab:CriticalTemper}
\end{center}
\vspace{-0.7cm}
\end{table}

\begin{figure}
\includegraphics[width=0.95\textwidth]{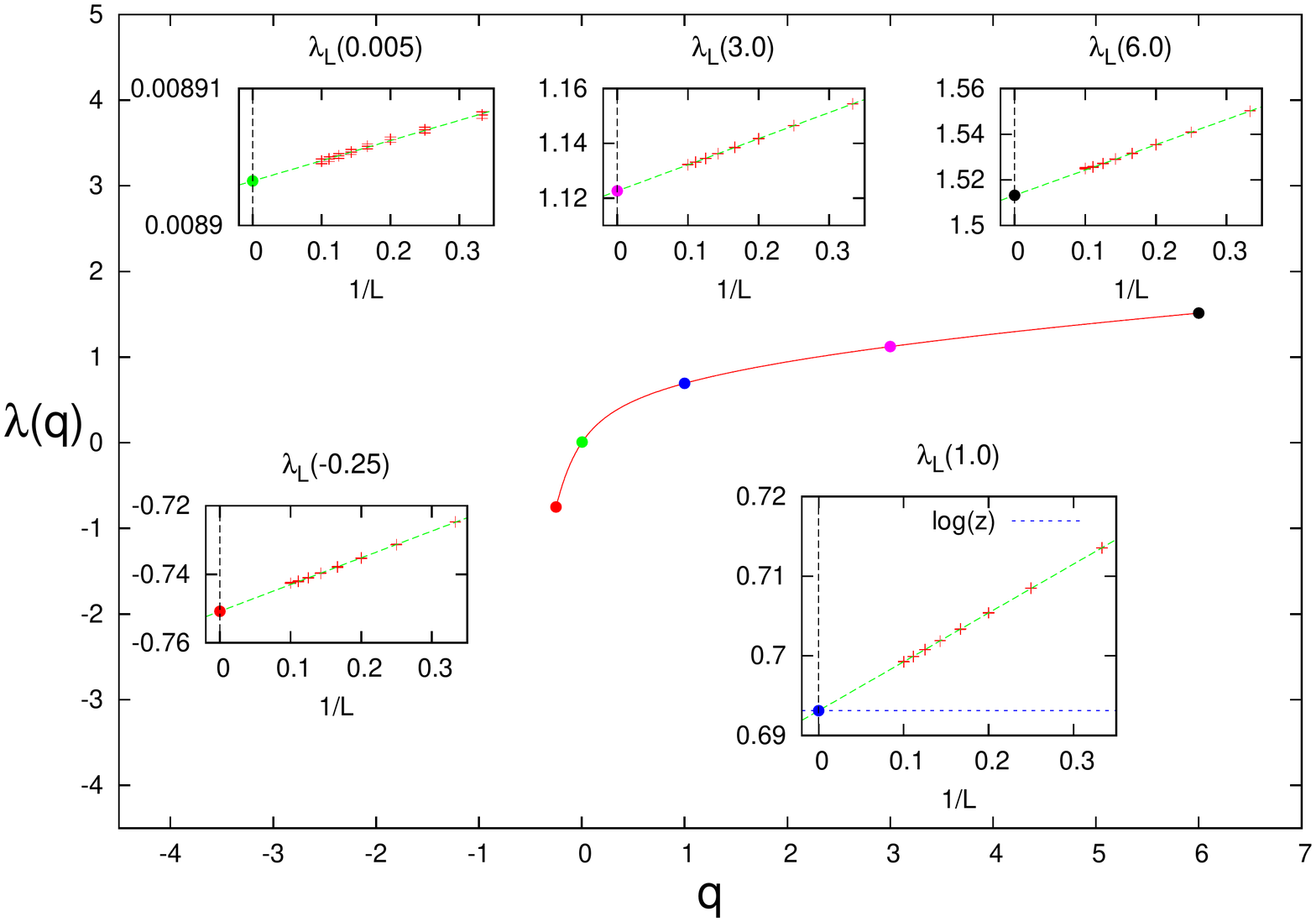}
\caption{Use of Method I to compute the cumulant generating function $\lambda(q)$. This plot refers to a function $\lambda(q)$ computed on a Bethe lattice with coordination number $z+1=3$, at the critical temperature corresponding to the random field strenght $\sigma=1.0$, which is approximatively $T_c(\sigma=1.0)\approx0.549$.
For the meaning of the insets, see the discussion in Section \ref{subsed:metodo1} .\label{fig:gamma} 
Notice that in most of the cases the errors are practically invisible.
(For typographical reason the letter $L$ stands for the $\ell$ used throughout the paper.)  }
\end{figure}

In \figurename\ref{fig:gamma} we show the function $\lambda(q)$ obtained with this method. The curve is computed at the critical temperature corresponding to the random field strenght $\sigma=1.0$ (and ferromagnetic coupling $J=1$),  for a Bethe lattice of connectivity $z+1=3$. We recall that the critical point is defined by the following condition:
\begin{equation}
\lambda(1)=\log(z)\ \ \ \text{for}\ \ \ T=T_c(\sigma).\label{eq:CondForTc}
\end{equation} 

The inset in  \figurename\ref{fig:gamma} corresponding to $q=1$ shows that, in our case, $\lambda(1)=\log(2)$. In general the condition \eqref{eq:CondForTc} defines a curve in the plane $(T,\sigma)$, separating the paramagnetic and the ferromagnetic phases. Some of the critical temperatures for the corresponding values of the random field $\sigma$  are sketched in  \tablename\ \ref{Tab:CriticalTemper}.
The other insets in \figurename\ref{fig:gamma} report the function $\lambda_{\ell}(q)$ for some representative values of $q$, together with the extrapolation to $\ell\rightarrow\infty$. 

Method I works quite well for values of $q$ in the range showed in \figurename\ref{fig:gamma}, i.e. for $0.25\leq q\leq6$. Outside this interval the error on the intermediate generating function \eqref{eq:lambdaintermediate} is quite large, so that the extrapolation $\ell\rightarrow\infty$ is far less reliable. This numerical shortcoming will be cured by the Method II.

\subsection{Method II}
Another route to work out  the function $\lambda(q)$, getting rid of the extrapolation $\ell\rightarrow\infty$, leads to the following eigenvalues equation\cite{Rizzo2014, Janzen}
\begin{equation}
\boxed{\mathbb{E}_r\int\dd u' g(u',q)\ \delta\left[u-\hat{u}(\beta, J, u'+ r)\right]\left(\frac{\partial\hat{u}}{\partial u'}\right)^q =e^{-\lambda(q)}g(u,q)}\  ,\label{eq:EqAutoval}
\end{equation}
where $\hat{u}(\beta, J, x)$ is given by Eq. \eqref{eq:uhat} and the average over the variable $r$ is performed using the distribution $P^{\text{cav}}_{z-1}(r)$ defined as:
\begin{equation}
P^{\text{cav}}_{z-1}(r)=\int\left[\prod_{i=1}^{z-1}\dd Q(u_i)\right]\overline{\delta\left(r-h_R-\sum_{i=1}^{z-1}u_i\right)}^{h_R}.
\end{equation}
Eq.\eqref{eq:EqAutoval}, for any fixed $q$, is a homogeneous Fredholm's integral equation of the second kind, which has a continous spectrum. Luckly, the function $\lambda(q)$ we are intersted in (i.e. the cumulant generating function of the decay rate) is obtained by solving Eq.\eqref{eq:EqAutoval}, for any fixed $q$, only for the largest eigevalue.\\ 
In order to derive Eq. \eqref{eq:EqAutoval} we start from the correlation function $\mathcal{C}(\ell)$:
\begin{equation}
\mathcal{C}(\ell) = \prod_{k=1}^{\ell}\frac{\partial u_{k\rightarrow k-1}}{\partial u_{k+1\rightarrow k}}\,\, .
\end{equation}
The random variables $u_{k\rightarrow k-1}$ and $u_{k+1\rightarrow k}$ are related by the following equation:
\begin{equation}
u_{k\rightarrow k-1} = \hat{u}(\beta,J,r+u_{k+1\rightarrow k})\ .
\end{equation}
The correlation function at distance $\ell+1$ can be written as:
\begin{equation}
\begin{aligned}
\mathcal{C}(\ell+1)&=\frac{\partial u_{1\rightarrow0}}{\partial u_{2\rightarrow1}}\mathcal{C}(\ell)\ ,\\
u_{1\rightarrow 0}&=\hat{u}(\beta,J,r+u_{2\rightarrow1})\ .\label{eq:recursionC}
\end{aligned}
\end{equation}
Let us consider the joint probability distribution of the variables $\mathcal{C}(\ell+1)$ and $u_{1\rightarrow0}$, which we call $P_{\ell+1}(\mathcal{C},u)$. Eq. \eqref{eq:recursionC} defines the following recursion relation for the function $P_{\ell}(\mathcal{C},u)$:
\begin{equation}
P_{\ell+1}(\mathcal{C},u)=\mathbb{E}_r\int \dd \mathcal{C}'\dd u'\ P_{\ell}(\mathcal{C}',u')\delta\left[\mathcal{C}-\frac{\partial \hat{u}(\beta,J,r+u')}{\partial u'}\mathcal{C}'\right]\delta[u-\hat{u}(\beta,J,r+u')]\ .\label{eq:recursionPC}
\end{equation}
Taking the partial moments: 
\begin{equation}
\psi_{\ell}(u,q)=\int \dd \mathcal{C}\ P_{\ell}(\mathcal{C},u)\ \mathcal{C}^q\ ,
\end{equation}
we find the following recursion for $\psi_{\ell}(u,q)$:
\begin{equation}
\psi_{\ell+1}(u,q)=\mathbb{E}_r\int \dd u'\ \psi_{\ell}(u',q)\ \delta[u-\hat{u}(\beta,J,r+u')]\left(\frac{\partial \hat{u}}{\partial u'}\right)^q\ .\label{eq:psi}
\end{equation}
For $\ell\rightarrow\infty$, Eq. \eqref{eq:psi} has only the solutions $\psi_{\infty}(u,q)\equiv 0$ for $q>0$ and $\psi_{\infty}(u,q)\equiv \infty$ for $q<0$, while for $q=0$ the asymptotic distribution is the cavity distribution $\psi_{\infty}(u,0)\equiv Q(u)$. Noticing that
\begin{equation}
\int \dd u\ \psi_{\ell}(u,q)=\overline{\mathcal{C}(\ell)^q}\ ,
\end{equation}
we define the renormalized function $g_{\ell}(u,q)$ as:
\begin{equation}
g_{\ell}(u,q)=\psi_{\ell}(u,q)\ e^{\ell\lambda(q)}\ ,\label{eq:g}
\end{equation} 
where $\lambda(q)$ is given by Eq. \eqref{eq:lambdaq}. The function $g_{\ell}(u,q)$ has a finite limit for $\ell\rightarrow\infty$ and an integral normalized to one:
\begin{equation}
\lim_{\ell\rightarrow\infty}\int\dd u\ g_{\ell}(u,q)=1.
\end{equation}
Inserting Eq. \eqref{eq:g} into Eq. \ref{eq:psi} and taking the limit $\ell\rightarrow\infty$, we recover Eq. \eqref{eq:EqAutoval}:
\begin{equation}
e^{-\lambda(q)}g(u,q)=\mathbb{E}_r\int \dd u'\ g(u',q)\ \delta[u-\hat{u}(\beta,J,r+u')]\left(\frac{\partial \hat{u}}{\partial u'}\right)^q\ . \nonumber
\end{equation}
The cumulant generating function is then given by: 
\begin{equation}
\boxed{
\lambda(q)=-\log\mathbb{E}_r\int \dd u\ g(u,q)\ \left(\frac{\partial \hat{u}}{\partial u}\right)^q\ .
}
\end{equation}

 Eq. \eqref{eq:EqAutoval} can be derived also with the replica method, by diagonalizing the disorder-averaged replicated transfer matrix:
\begin{equation}
 T_n(s,s')\ =\ \mathbb{E}_r\ \exp\left(\beta J\sum_{a=1}^ns_as'_a+\beta r\sum_{a=1}^n s'_a\right)\ ,\label{eq:TransfMatx}
\end{equation}
 and taking the limit $n\rightarrow0$ in the corresponding eigenvalues equations\cite{Weigt1996}.

It is worth recalling how the replica method works and how the physical properties are described in that language. So, we briefly summarize the analysis made in Ref.~\onlinecite{Weigt1996}, bridging the replica approach with the cavity interpretation.   
 The diagonalization of the replicated transfer matrix \eqref{eq:TransfMatx} is well defined only for integer values of the number of replicas $n$. Exploiting the symmetry of $T_n(\sigma,\sigma')$ under replica permutations, it can be block-diagonalized according to the irreducibile representations (IR) of the symmetric group. Each IR $D^{(q,m)}$ is labeled by two integer numbers $q=0,\dots\lfloor n/2\rfloor$ and $m=1,\dots,(n+1-2q)$. For any given value of $q$, the $(n+1-2q)$ IR $D^{(q,m)}$ are all isomorphic and have the same dimension $d_q=\text{dim}(D^{(q,m)})=\binom{n}{q}-\binom{n}{q-1}$. The replicated transfer matrix $T_n^{(q)}$, restricted to the q-sector, will contain $(n+1-2q)$ different eigenvalues $\mu^{(q)}_m$, each one with degeneracy $\text{deg}\left(\mu^{(q)}_m\right)=d_q$. A graphic visualization of the q-sector of $T_n$ is the following:
\begin{equation}
\begin{array}{c}
T_{n}^{(q)}=\underbrace{\left[\begin{array}{ccc}
\begin{array}{c}
D^{(q,1)}=\underbrace{\left(\begin{array}{ccc}
\mu_{1}^{(q)} & 0 & 0\\
0 & \ddots & 0\\
0 & 0 & \mu_{1}^{(q)}\end{array}\right)}\\
\,\,\,\,\,\,\,\,\,\,\,\,\,\,\,\,\,\,\,\,\,\,\, d_{q}\end{array} &  & 0\\
 & \ddots\\
0 &  & \begin{array}{c}
D^{(q,n+1-2q)}=\underbrace{\left(\begin{array}{ccc}
\mu_{n+1-2q}^{(q)} & 0 & 0\\
0 & \ddots & 0\\
0 & 0 & \mu_{n+1-2q}^{(q)}\end{array}\right)}\\
\,\,\,\,\,\,\,\,\,\,\,\,\,\,\,\,\,\,\,\,\,\,\,\,\,\,\,\,\,\,\,\,\,\,\,\,\,\,\, d_{q}\end{array}\end{array}\right]}\\
d_{q}\times(n+1-2q)
\end{array}\ .
\end{equation}
The sector $q=0$ contains the replica-simmetric IR $D^{(0,m)}$, with $m=1,\dots,n+1$. The IR $D^{(0,m)}$ are unidimensional representations and each one carries a different eigenvalue $\mu_m^{(0)}$. The maximum eigenvalue $\mu^{(0)}=\text{max}_m\left(\mu_m^{(0)}\right)$ is equal to 1 and it is related to $\lambda(0)$ by the equation:
\begin{equation}
\mu^{(0)}=e^{-\lambda(0)}\ .
\end{equation}
In  the replica language the sector $q=0$ is called the \textit{longitudinal} sector. Sectors with $q\neq0$ correspond to replica-symmetry broken representations. Calling $\mu^{(q)}$ the largest eigenvalue belonging to the $q$-sector, $\mu^{(q)}=\text{max}_m(\mu_m^{(q)})$, we have the following identification:
\begin{equation}
\mu^{(q)}=e^{-\lambda(q)}\ .\label{replicaID}
\end{equation}

The sector with $q=1$ is called the \textit{anomalous} sector and that with $q=2$ is the \textit{replicon} sector. The largest eigenvalue of the anomalous sector controls the decay of the average connected correlation function: $\overline{\mathcal{C}(\ell)}=(\mu^{(1)})^{\ell}$, thus giving the signature of the ferromagnetic instability when it assumes the value $\mu^{(1)}=z^{-1}$. The largest eigenvalue of the replicon sector controls the decay of the average squared connected correlation function: $\overline{\mathcal{C}(\ell)^2}=(\mu^{(2)})^{\ell}$ and signals the spin-glass instability when it evaluates $\mu^{(2)}=z^{-1}$. In general it can be proven\cite{MP, Weigt1996} that the largest eigenvalue of each sector $q$ controls the decay of the $q^{th}$-moment of the connected correlation function: $\overline{\mathcal{C}(\ell)^q}=(\mu^{(q)})^{\ell}$. Moreover all the contribution to the assessment of the moment $\overline{\mathcal{C}(\ell)^q}$ comes only from the sector $q$, since other sectors with $k\neq q$ give vanishing contributions. Anyhow the sectors with $q\geq3$ do not have a specific nomenclature and they are not associated to any kind of phase transition (that is $\mu^{(q)}<z^{-1}$ for $q\geq3$). 

At this stage the replica results are only justified for integer values of $q$. Taking the limit $n\rightarrow0$, the analytic continuation to real values of $q$ allows one to obtain a continous set of eigenvalues $\mu^{(q)}$.

In the replica picture the cumulant generating function $\lambda(q)$ is related to the spectrum of the replicated transfer matrix via Eq. \eqref{replicaID}. The eigenfunctions $g(u,q)$ correspond to eigenperturbations towards replica-symmetry broken directions.

In many applications the knowledge of the spectrum of the replicated average transfer matrix has a deep theoretical importance, particularly in the computation of the finite size $1/N$ corrections in finitely connected models, see for example Refs.~\onlinecite{MP,Mio}.

Method II looks much more robust than Method I, mainly because it circumvents the annoying problem of extrapolating the $\ell\rightarrow\infty$ limit. A way to solve  numerically the eigenvalues equation \eqref{eq:EqAutoval} requires discretizing the variables and replacing integral by sum. If the integral kernel is positive defined (as in the case of the RFIM) it can be solved also by means of a population dynamics algorithm. In adopting the latter scheme one has to distinguish the algorithm for $q>0$ from that for $q<0$. If $q>0$ the population is reweighted by the factor $\left(\frac{\partial \hat{u}}{\partial u'}\right)^q<1$. Practically, when running the population dynamics algorithm, the updates are accepted with probability $\left(\frac{\partial \hat{u}}{\partial u'}\right)^q$. Instead if $q<0$ the same factor is bigger than one $\left(\frac{\partial \hat{u}}{\partial u'}\right)^q>1$. The effect on the algorithm is that of introducing in the population, at each update, more copies of the same elements. This requires to filter the population at each step of the algorithm, in order to reduce the fraction of twins from the pool of evolving fields representing the population. 

\begin{figure}
\includegraphics[width=0.75\textwidth]{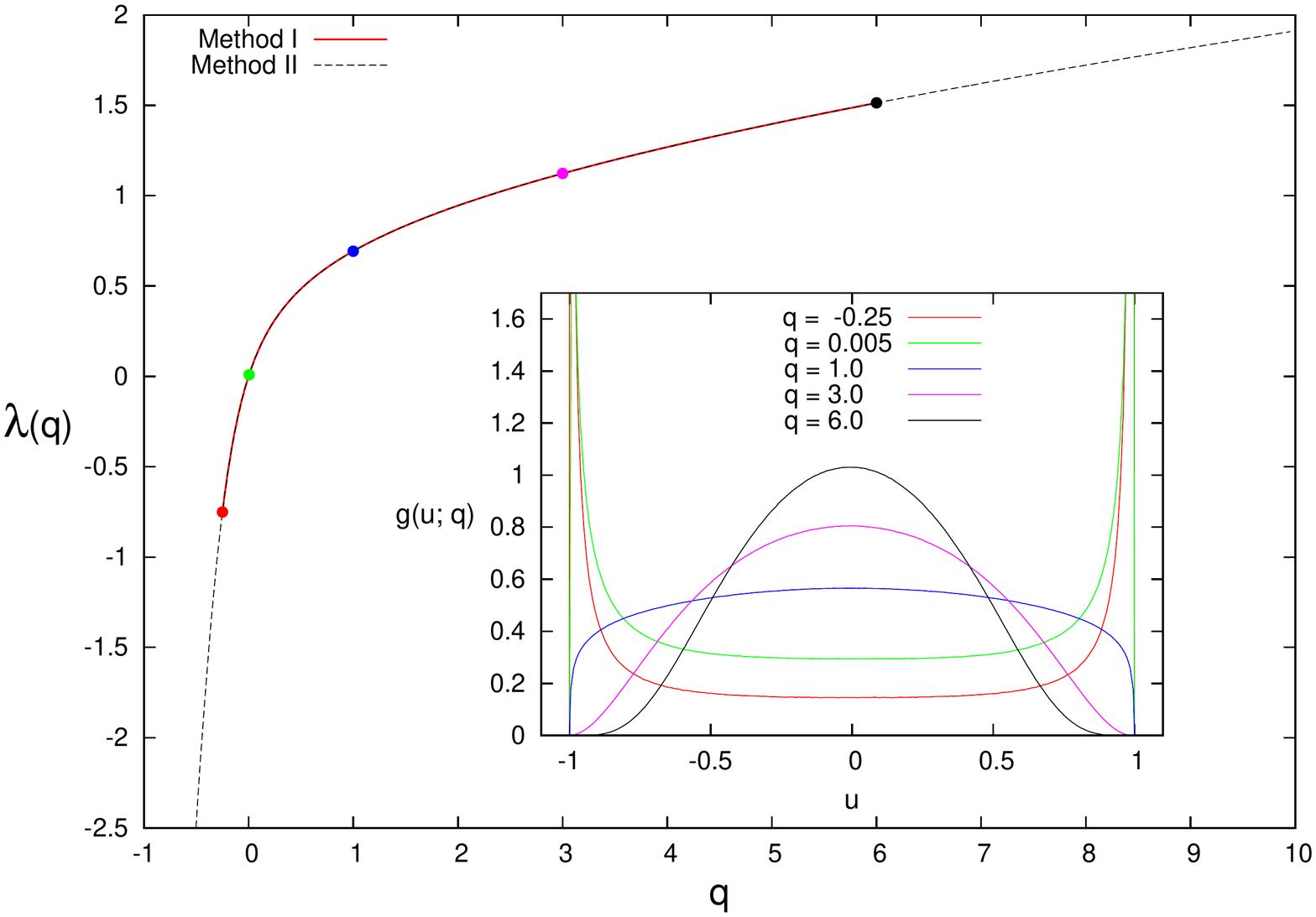}
\caption{The cumulant generating function $\lambda(q)$ computed with Method I (red line) and Method II (black line) on a Bethe lattice with fixed coordination number $z+1=3$, at the critical point corresponding to the random field strength $\sigma=1.0$ and temperature $T_c=0.549$. In the inset we show the eigenfunctions $g(u,q)$ computed for the values of $q$ marked by dots in the main panel.}
\label{fig:compareLambda}
\end{figure}

In \figurename\ref{fig:compareLambda} we compare the function $\lambda(q)$ computed with Method I and Method II. In the same figure we also show some of the eigenfunctions $g(u,q)$. 

\begin{figure}
\includegraphics[width=0.8\textwidth]{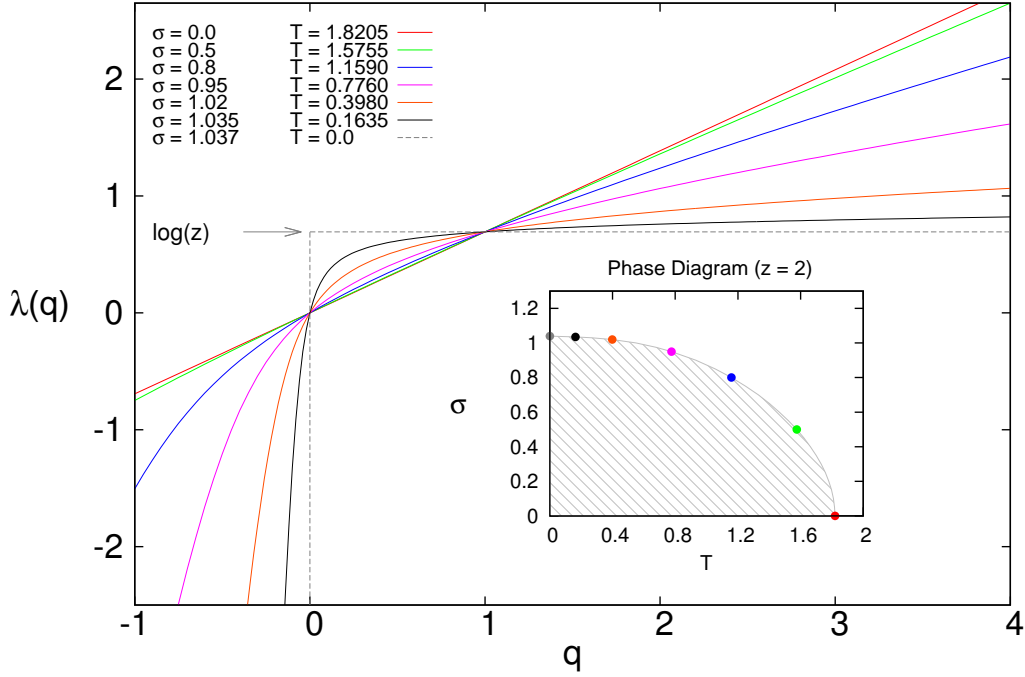}
\caption{Profiles of the cumulant generating function $\lambda(q)$ computed at different critical points along the critical line, shown in the inset. In zero field $\lambda(q)$ assumes the simple form $\lambda(q)=q\log(z)$. At zero temperature $\lambda(q)$ is piecewise constant: $\lambda(q)=-\infty$ for $q<0$, $\lambda(q)=0$ for $q=0$ and $\lambda(q)=\log(z)$ for $q>0$.}
\label{fig:lambdaTc}
\end{figure}

In  \figurename\ref{fig:lambdaTc} we draw the profiles of the function $\lambda(q)$ at different critical points on the critical line, ranging from the zero field critical point to the zero temperature critical point.

 In the zero field case the function $\lambda(q)$ is simply given by $\lambda(q)=q\log(z)$. At the zero temperature critical point the function $\lambda(q)$ becomes singular. This is due to the fact that Eq. \eqref{eq:EqAutoval}, in the limit $\beta\rightarrow\infty$, becomes:
\begin{equation}
e^{-\lambda(q)}g(u,q)=\mathbb{E}_r\int \dd u'\ g(u',q)\ \delta[u-\hat{u}(J,r+u')]\Theta(J-|r+u'|)^q\ ,
\end{equation}
where $\hat{u}(J,r+u')=\text{sign}(r+u')\min(J,|r+u'|)$. We deduce that for any $q>0$ the function $\lambda(q)$ must be a costant. At the zero temperature critical point it remains true that $\lambda(1)=\log(z)$, so that $\lambda(q)=\log(z)$ for any $q>0$. For $q=0$ the eigenvalues equation reduces to cavity equation at zero temperature and we find $\lambda(q=0)=0$. Lastly for $q<0$ we have $\lambda(q)=-\infty$.

A further comment on the limits $q\rightarrow\infty$ and $q\rightarrow-\infty$ can be useful.
In the $q \to \infty$ limit, the average $\overline{\mathcal{C}(\ell)^q}$ is dominated by the largest correlations $\mathcal{C}(\ell)$. By noticing that each term in the product in Eq.(\ref{eq:Cnorm}) is bounded by
\[
\frac{\partial u_{k \to k-1}}{\partial u_{k+1\to k}} \le \tanh(\beta J)\;,
\]
we have that the largest correlations approach the upper bound
\[
\mathcal{C}(\ell)^q \le \tanh(\beta J)^{q\ell}\;,
\]
that in turn implies
\[
\lambda(q) \ge -q \log\tanh(\beta J)\;.
\]
Since the function $\lambda(q)$ is concave, it can not be super-linear and so, for large $q$ values, it must grow linearly
\[
\lambda(q) \approx - q \log\tanh(\beta J)\;.
\]

On the contrary, the $q\to -\infty$ limit depends on the distribution of the random fields: if random fields are bounded then a linear behavior in $\lambda(q)$ can be derived with an argument similar to the one above, but if random fields are unbounded (as in the case of Gaussian random fields) the behavior of $\lambda(q)$ for $q\to -\infty$ depends on the tail of the distribution of the random fields.
Indeed a correlation can become very close to zero only if a large random field is generated, since
\[
\frac{\partial u_{k \to k-1}}{\partial u_{k+1\to k}} \approx 2 \sinh(2 \beta J) e^{-\beta |h_R|}\qquad\text{for } |h_R|\gg 1\;.
\]
For Gaussian random fields the leading behaviour for $q \to -\infty$ is
\[
\lambda(q) \approx -2 \beta^2 \sigma^2 q^2\;.
\]

\section{Computing the rate function $\Sigma(\gamma)$}
The rate function $\Sigma(\gamma)$ is the Legendre-Fenchel trasform of the function $\lambda(q)$ via Eq.\eqref{eq:ratefunction}. To estimate $\Sigma(\gamma)$ from $\lambda(q)$ we use the following routine. The input data are the pairs $\{q_i,\lambda(q_i)\}_{i=1}^{N_q}$, where $N_q$ is the number of available values of $q$. We start by computing the local slopes $s_i$:
\begin{equation}
s_i = \frac{\lambda(q_{i+1})-\lambda(q_i)}{q_{i+1}-q_i}\ .
\end{equation}
Since the function $\lambda(q)$ is concave, the sequence is only decreasing.
Then we choose $N_\gamma$ values $\gamma_1,\dots,\gamma_{N_\gamma}$, and for each $\gamma_k$ we obtain $\Sigma(\gamma_k)$ as:
\begin{equation} 
\text{If}\ \ \  s_{i}< \gamma_k \leq s_{i-1}\ \ \ \text{then}\ \ \ \Sigma(\gamma_k)=\lambda(q_{i-1})-q_{i-1}\gamma_k\ .
\end{equation}
We notice that $N_q=N_\gamma$ is required to obtain good numerical accuracy.

Let us recall two properties concerning the rate function $\Sigma(\gamma)$.
The point where $\Sigma(\gamma)=0$ identifies the typical value $\gamma=\gamma_0$ of the decay rate. This quantity is rather easy to compute, since from the definition \eqref{eq:lambdaq} we have:
\begin{equation}
\gamma_0=\frac{\dd\lambda(q)}{\dd q}\Bigg|_{q=0}=-\lim_{\ell\rightarrow\infty}\frac{\overline{\log\mathcal{C}(\ell)}}{\ell}=-\mathbb{E}_{h}\log\left[\frac{\partial}{\partial h}\hat{u}(\beta,J,h)\right],\label{eq:gamma0}
\end{equation}
where $\hat{u}(\beta, J, x)$ is given by Eq. \eqref{eq:uhat} and the expectation over the cavity field $h$ is taken using the distribution \eqref{eq:Pcav}.

The point $\gamma^*$, defined by the equation:
\begin{equation}
\frac{\partial\Sigma(\gamma)}{\partial\gamma}\Big|_{\gamma^*}=-1\ ,
\end{equation} 
 gives the rate of correlations which dominate the average $\overline{\mathcal{C}(\ell)}$. The value of $\Sigma(\gamma)$ in this point is: $\Sigma(\gamma^*)=\lambda(1)-\gamma^*$. At the critical point we know from Eq. \eqref{eq:CondForTc} that $\lambda(1)=\log(z)$, and then we have:
\begin{equation}
\Sigma(\gamma^*)=\log(z)-\gamma^*\ \ \ \text{for}\ \ \ T=T_c(\sigma)\,\, .
\end{equation}

\begin{figure}
\centering
{\includegraphics[width=0.8\textwidth]{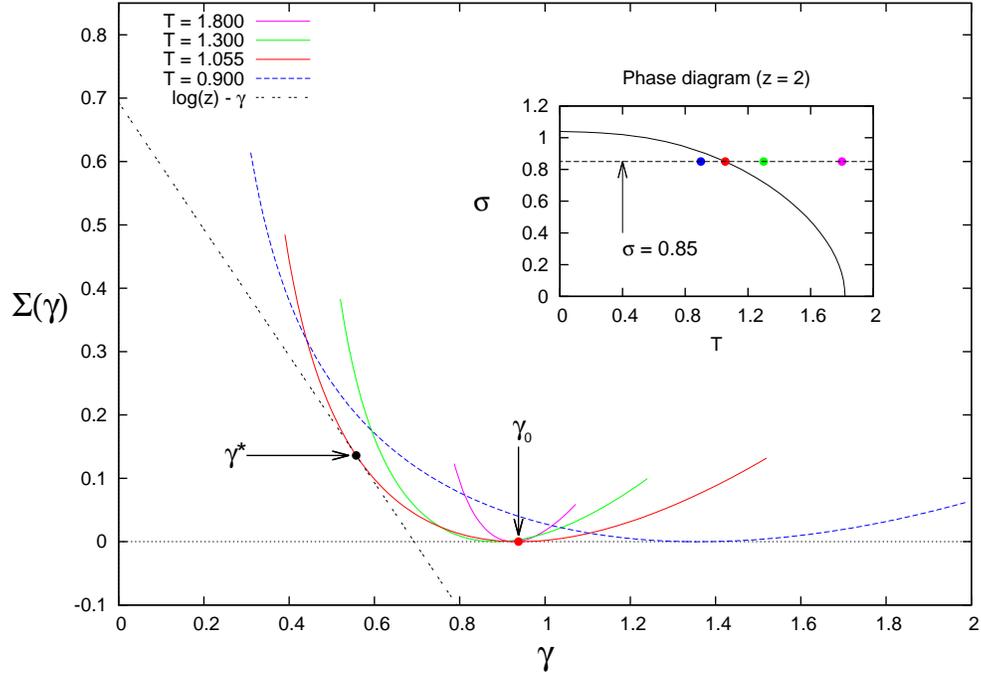}}
\caption{ Rate function $\Sigma(\gamma)$, computed at fixed random field strenght $\sigma=0.85$ and four different temperatures: $T = 1.8$ (purple line), $T = 1.3$ (green line), $T = 1.055=T_c$ (red line), $T=0.9$ (blue line). The point $\gamma=\gamma_0$ is the solution of the equation $\Sigma(\gamma)=0$ and represents the typical value of the decay rate. The point $\gamma=\gamma^*$ is the solution of the equation $\Sigma'(\gamma)+1=0$ and represents the decay rate of correlations contributing the most to the assessment of $\overline{\mathcal{C}(\ell)}$.
\label{fig:Sigma0.85}}
\end{figure}

In  \figurename\ref{fig:Sigma0.85} we show the profiles of the rate function $\Sigma(\gamma)$, computed on a Bethe lattice of connectivity $z+1=3$, at a fixed random field strenght equal to $\sigma = 0.85$  and four temperatures: $T = 1.8>T_c$, $T = 1.3>T_c$, $T = 1.055=T_c$, $T=0.9<T_c$. This gives an idea on how the function $\Sigma(\gamma)$ changes when lowering the temperature. Two opposite effects should be noted: close to $T_c$, $\gamma_0$ grows while the curvature of the function decreases. The critical temperature $T_c$ is the only temperature such that the curve $\Sigma(\gamma)$ touches the straight line $\log(z)-\gamma$.

\begin{figure}
\includegraphics[width=0.6\textwidth]{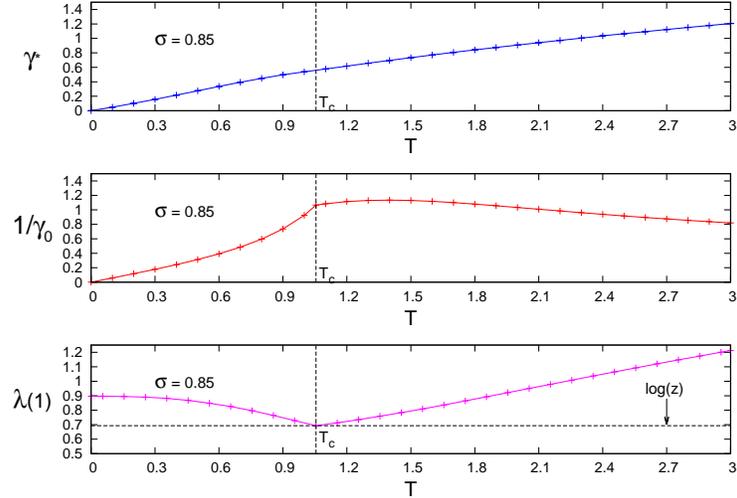}
\caption{ Behaviours of the rate $\gamma^*$ (upper panel), the inverse of the typical rate $1/\gamma_0$ (central panel) and the decay rate of the average correlation function $\lambda(1)$ with the temperature $T$ for a fixed random field strenght equals to $\sigma=0.85$. The rates $\gamma^*$ and $\lambda(1)$ are related by the equation $\Sigma(\gamma^*)+\gamma^*=\lambda(1)$.\label{fig:rates0.85}
}
\end{figure}

The behaviours of the quantities $\gamma^*$, $\gamma^0$ and $\lambda(1)$ as functions of the temperature, at fixed random field strenght $\sigma=0.85$, are depicted in \figurename\ref{fig:rates0.85}. We recall that the rate $\gamma^*$ and the rate $\lambda(1)$ (i.e. the decay rate of the average correlation function: $\overline{\mathcal{C}(\ell)}=e^{-\ell\lambda(1)}$) are related by the equation $\lambda(1)=\gamma^*+\Sigma(\gamma^*)$.

It is interesting to see how the rate function changes when moving along the critical line. To this end we consider the total number of correlations $\mathcal{C}(\ell)$ decaying with rate $\gamma$:
\begin{equation}
\mathcal{N}_{\ell}(\gamma)=z^{\ell}e^{-\ell\Sigma(\gamma)}\ ,
\end{equation}
and we define the logarithmic scaled number as:
\begin{equation}
n(\gamma)=\lim_{\ell\rightarrow\infty}\frac{\log\mathcal{N}_{\ell}(\gamma)}{\ell}=\log(z)-\Sigma(\gamma)\ .
\end{equation}
The word "scaled" is used to indicate that $n(\gamma)$ is the logarithm of $\mathcal{N}_{\ell}(\gamma)$ \textit{scaled} by $\ell$.
From the definition of $n(\gamma)$ we will have that $n(\gamma_0)=\log(z)$ and $n(\gamma^*)=\gamma^*$. 

\begin{figure}
\includegraphics[width=0.8\textwidth]{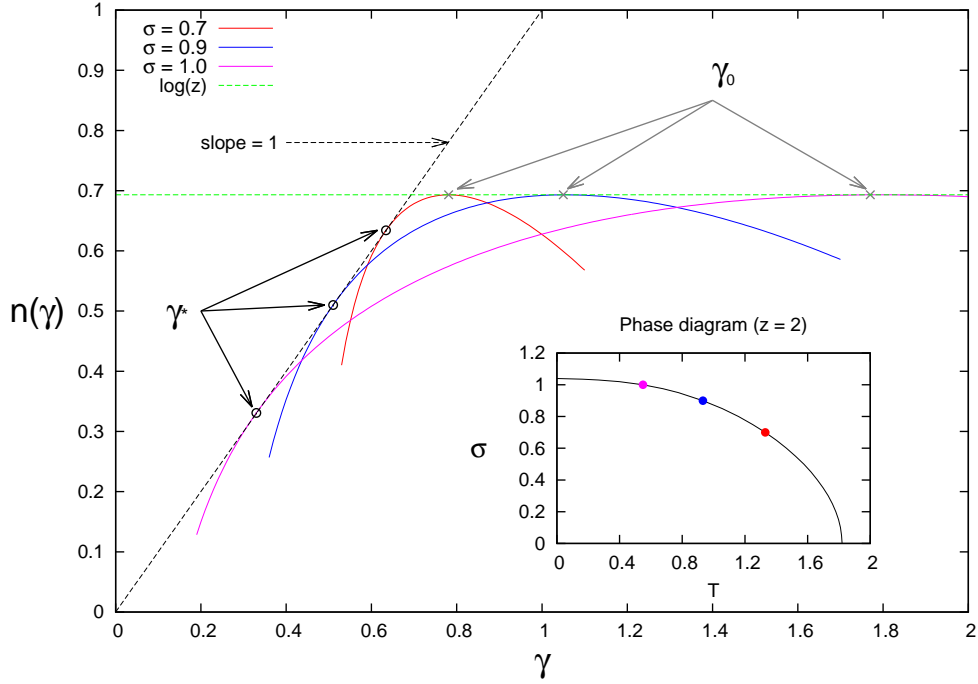} 
\caption{Profiles of the logarithmic scaled number $n(\gamma)$ of correlations with decay rate $\gamma$ in a graph of connectivity $z+1=3$. The curves were computed on the critical line for several values of the random field strenght. Crosses are solutions of the equation $\Sigma(\gamma)=0$ and represent the typical correlations in the system. Circles represent correlations contributing the most to the average $\overline{\mathcal{C}(\ell)}$ and they are obtained as solutions of the equation $\Sigma^{\prime}(\gamma)=-1$.\label{fig:Ldf} }
\end{figure}

In \figurename\ref{fig:Ldf} we show the logarithmic scaled number $n(\gamma)$ computed in three different critical points along the critical line. 

\begin{figure}
\includegraphics[width=0.6\textwidth]{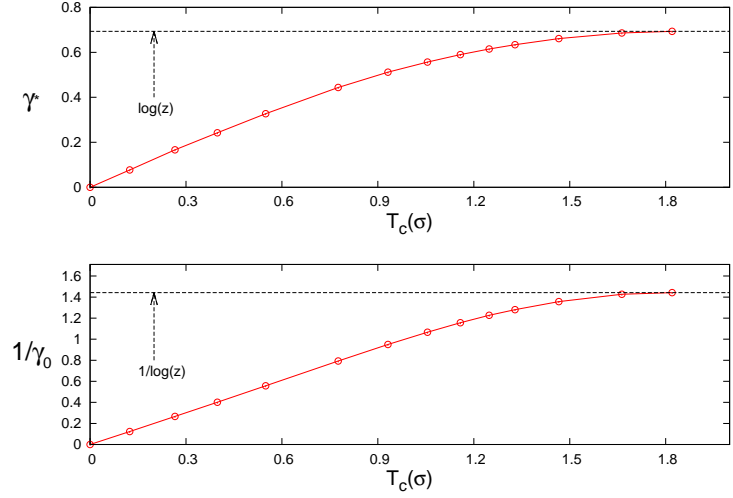} 
\caption{The rate $\gamma^*$ (upper panel) and the inverse of the typical rate $1/\gamma_0$ (lower panel) as functions of the critical temperature $T_c(\sigma)$. \label{fig:gammas}}
\end{figure}

In \figurename\ref{fig:gammas} we report the behaviour of the saddle point decay rate $\gamma^*$ and the typical decay rate $\gamma_0$, as functions of the critical temperature $T_c(\sigma)$. 

We note that, moving toward the zero temperature critical point, the typical rate $\gamma_0$ becomes bigger and bigger and it can be seen, from Eq. \eqref{eq:gamma0}, that it diverges as $T^{-1}$ for $T\rightarrow 0$.
Indeed in the limit $T\rightarrow0$ Eq. \eqref{eq:gamma0} becomes:
\begin{equation}
\gamma_0 \simeq \int \dd h P(h)\ \log[1+e^{2\beta(|h|-J)}]\simeq\int_{|h|<J}\dd h P(h)\ e^{-2\beta(J-|h|)}
+2\beta\int_{|h|>J}\dd h P(h)(|h|-J)\propto \frac{1}{T}\ \ \ \text{for}\ \ \ T\rightarrow 0\ .
\end{equation}
 On the contrary the rate $\gamma^*$ becomes smaller and smaller and it is expected to vanish for $T\rightarrow 0$. This means that stronger and stronger correlations dominate the critical behaviour when going to low temperatures. By virtue of the fact that $n(\gamma^*)=\gamma^*$ this also means that dominant correlations becomes even more rare.
This observation suggests that, exactly at zero temperature, the critical behaviour is induced by a sub-exponential number of correlated spin variables. 
This number can be estimated in the following way.
 Let us consider a chain of spins of lenght $\ell$. The energy of the chain conditioned on the boundary spins, apart from an additive costant, can be written as:
\begin{equation}
E_{\ell}(s_0,s_{\ell})\ =\ -h_0s_0-h_{\ell}s_{\ell}-J_{\ell}s_0s_{\ell}\ .
\end{equation}
Knowing the triplet $(h_0,h_{\ell},J_{\ell})$, we can compute the magnetization $\langle s_0\rangle$:
\begin{equation}
\langle s_0\rangle = \tanh\left\{\beta h_0+\text{atanh}[\tanh(\beta J_{\ell})\tanh(\beta h_{\ell})]\right\}\ .
\end{equation}
From $\langle s_0\rangle$ we obtain the response function $\mathcal{R}(\ell;h_0,h_{\ell},J_{\ell})$:
\begin{equation}
\mathcal{R}(\ell;h_0,h_{\ell},J_{\ell})=\frac{\partial\langle s_0\rangle}{\partial h_{\ell}}=\beta(1-\langle s_0\rangle^2)\frac{\text{th}(\beta J_{\ell})(1-\text{th}^2(\beta h_{\ell}))}{1-\text{th}^2(\beta J_{\ell})\text{th}^2(\beta h_{\ell})}\ ,
\end{equation}
and the normalized correlation function $\mathcal{C}(\ell;h_{\ell},J_{\ell})$:
\begin{equation}
\mathcal{C}(\ell;h_{\ell},J_{\ell})=\frac{\text{th}(\beta J_{\ell})(1-\text{th}^2(\beta h_{\ell}))}{1-\text{th}^2(\beta J_{\ell})\text{th}^2(\beta h_{\ell})}\ .
\end{equation}
In the limit $\beta\rightarrow\infty$, the correlation $\mathcal{C}(\ell;h_{\ell},J_{\ell})$ becomes a theta function:
\begin{equation}
\mathcal{C}(\ell;h_{\ell},J_{\ell})=\Theta(J_{\ell}-|h_{\ell}|)\ .
\end{equation}
Now, let us take the average of  $\mathcal{C}(\ell;h_{\ell},J_{\ell})$ with respect to $h_{\ell}$:
\begin{equation}
\mathcal{C}(\ell;J_{\ell})=\int \dd h\  P_{\ell}(h)\  \Theta(J_{\ell}-|h|)\ .
\end{equation}
For $\ell$ large enough we can assume $J_{\ell}\ll 1$ and we find:
\begin{equation}
\mathcal{C}(\ell;J_{\ell})\propto J_{\ell}\ , \ \ \text{for}\ \ \ \ell\rightarrow\infty\ .
\end{equation}
This means that all non-zero correlations still have a non-trivial distribution, which can be asymptotically inferred from the distribution of the effective coupling between two spins. If we get this distribution, we can find the number of critical correlations with sub-exponential accuracy.  
This is precisely the goal of the next Section.

\section{Large Deviations at Zero Temperature}\label{sec:ZT}

In the limit $T\rightarrow 0$ the normalized connected correlation function $\mathcal{C}(\ell)$ can assume only two values, namely $0$ and $1$, so that the cumulant generating function $\lambda(q)$ becomes not differentiable. If we imagine to go to lower and lower temperatures, we would find, on one side, that typical correlations become closer and closer to zero and the typical rate $\gamma_0$ would be pushed to $+\infty$. On the other side correlations which dominate the susceptibility would become closer  and closer to $1$ and the rate $\gamma^*$ would tend to $0$. At the same time the number $\mathcal{N}_{\ell}(\gamma^*)$ of atypical correlations becomes sub-exponential in the limit $T\rightarrow 0$, so that large deviations on the exponential scale disappear.

The really interesting observable at $T=0$ is the response function $\mathcal{R}(\ell)$, which is the quantity directly involved in the computation of the susceptibility $\chi_F$. In this Section we derive the analytical form of  the response function distribution. To this end we again consider two spins in the graph, named $s_0$ and $s_{\ell}$, joined by a chain of lenght $\ell$. Calling $h_0$ and $h_{\ell}$ the local fields acting on $s_0$ and $s_{\ell}$, and $J_{\ell}$ their effective coupling, the state of the spin $s_0$ in the ground state is 
\begin{equation}
\sigma_{0}^{\mathrm{GS}}=\Theta(J_{\ell}-|h_0|)\mathrm{sgn}(h_0+h_{\ell})+\Theta(|h_0|-J_{\ell})\mathrm{sgn}(h_0)\,\, ,
\end{equation}
and the response function is given by 
\begin{equation}
\mathcal{R}(\ell;h_0,h_{\ell},J_{\ell})=\frac{\partial\sigma_{0}^{\mathrm{GS}}}{\partial h_{\ell}}=2\delta(h_0+h_{\ell})\Theta(J_{\ell}-|h_0|)\,\, .
\end{equation} 
Calling $\mathcal{P}(h_0,h_{\ell})$ the joint density of $h_0$ and $h_{\ell}$ and taking the average of $\mathcal{R}(\ell)$ we have
\begin{equation}
\mathcal{R}(\ell;J_{\ell})=2\int dh\mathcal{P}(h,-h)\Theta(J_{\ell}-|h|)\,\, ,\label{eq:av_resp}
\end{equation}
When $\ell\rightarrow\infty$, the effective coupling is very small $J_{\ell}\ll 1$ and eq \eqref{eq:av_resp} can be approximated as 
\begin{equation}
\mathcal{R}(\ell;J_{\ell})\approx4\mathcal{P}(0,0)J_{\ell}\,\, .
\end{equation}
The response function, averaged over the boundary fields, becomes proportional to the effective coupling between the variables, for large distances. The ferromagnetic susceptibility is 
\begin{equation}
\chi_F = \sum_{\ell}z^{\ell}\ \overline{\mathcal{R}(\ell)}\propto\sum_{\ell}z^{\ell}\ \overline{J_\ell}\,\, .
\end{equation}
The information about rare correlations is contained in the pdf $P_{\ell}(J)$. To compute the function $P_{\ell}(J)$, we consider the energy of a chain in the graph connecting two spins $s_0$ and $s_{\ell}$, conditioned on the boundary spins:
\begin{equation}
E_{\ell}(s_0,s_{\ell})=-h_0^{(\ell)}s_0-h_{\ell}s_{\ell}-J_{\ell}s_0s_{\ell}+\mathcal{E}_{\ell}\,\, ,
\end{equation}
where $\mathcal{E}_{\ell}$ is a costant not depending on the boundary spins. Adding a new spin at the end of the path $s_{\ell+1}$ and minimizing the energy over $s_{\ell}$, one gets the new energy function\footnote{We assume in this section, without any loss of generality, the ferromagnetic interaction strenght to be $J=1$.} $E_{\ell+1}(s_0,s_{\ell+1})$:
\begin{equation}
E_{\ell+1}(s_0,s_{\ell+1}) = -h_0^{(\ell+1)}s_0-h_{\ell+1}s_{\ell+1}-J_{\ell+1}s_0s_{\ell+1}+\mathcal{E}_{\ell+1}\,\, ,
\end{equation}
with
\begin{align}
h_{0}^{(\ell+1)} &= h_{0}^{(\ell)}+\mathrm{sgn}(h_{\ell})\Delta h_0^{(\ell+1)}\,\, ,\\
h_{\ell+1} &= r+\mathrm{sgn}(h_{\ell})\Delta h_{\ell+1}\,\, ,\\
\mathcal{E}_{\ell+1}&=\mathcal{E}_{\ell} + \Delta_{\mathcal{E}}^{(\ell+1)}\,\, ,
\end{align}
where $r$ is the cavity field coming from the $z-1$ branches outside the path that merge on the node $\ell+1$.
The evolution rules for $\Delta h_0^{(\ell)}$, $\Delta h_{\ell}$, $J_{\ell}$, $ \Delta_{\mathcal{E}}^{(\ell)}$ are reported in \tablename\ \ref{Tab:FlowZt}.
\begin{table}[h]
\begin{center}
\begin{tabular}{|l|c|c|c|c|}
\hline
& $\Delta h_0^{(\ell+1)}$ & $\Delta h_{\ell+1}$ & $J_{\ell+1}$ & $ \Delta_{\mathcal{E}}^{(\ell+1)}$\\
\hline
I)\ \ \ \ \ \ \ \ $|h_{\ell}|<1-J_{\ell}$ & $0$ & $|h_{\ell}|$ & $J_{\ell}$ & $-1$ \\
\hline
II)\ \ \ $1-J_{\ell}<|h_{\ell}|<1+J_{\ell}\ \ $ & $\ \ (|h_{\ell}|+J_{\ell}-1)/2\ \ $ & $\ \ (|h_{\ell}|+1-J_{\ell})/2\ \ $ & $\ \ (1+J_{\ell}-|h_{\ell}|)/2\ \ $ & $\ \ -(1+|h_{\ell}|+J_{\ell})/2\ \ $ \\
\hline
III)\ \ \ \ \ \ $|h_{\ell}|>1+J_{\ell}$ & $J_{\ell}$ & $1$ & $0$ & $-|h_{\ell}|$ \\
\hline
\end{tabular}
\caption{Flow of the parameters of the boundary-spins-conditioned energy function $E_{\ell}(\sigma_0,\sigma_{\ell})$ in the evolution $E_{\ell}(\sigma_0,\sigma_{\ell})\rightarrow E_{\ell+1}(\sigma_0,\sigma_{\ell+1})$}\label{Tab:FlowZt}
\end{center}
\vspace{-0.7cm}
\end{table}
\\\\
The recursive equation for the joint distribution of $h_{\ell}$ and $J_{\ell}$ is:
\begin{align}
P_{\ell+1}(J,h)=\mathbb{E}_r\int_0^1 \dd J^{\prime}&\Big\{\int_{I}\dd h'\ P_{\ell}(J^{\prime},h^{\prime})\delta(J-J^{\prime})\delta(h-r-h^{\prime})\nonumber\\
&+\int_{II}\dd h'\ P_{\ell}(J^{\prime},h^{\prime})\delta\Big(J-\frac{1+J^{\prime}-|h^{\prime}|}{2}\Big)\delta\Big(h-r-\mathrm{sgn}(h^{\prime})\frac{|h^{\prime}|+1-J^{\prime}}{2}\Big)\nonumber\\
&+\int_{III}\dd h'\ P_{\ell}(J^{\prime},h^{\prime})\delta(J)\delta(h-r-\mathrm{sgn}(h^{\prime}))\Big\}\,\, \label{eq:recursionP},
\end{align}
where $\mathbb{E}_r$ is the expectation over the field $r$, which is distributed according to $P^{\mathrm{cav}}_{z-1}(r)$. The integration domains I, II and III are defined in \tablename\ \ref{Tab:FlowZt}\ . Eq. \eqref{eq:recursionP} has to be solved with the initial condition $P_1(J,h)=\delta(J-1)P^{\mathrm{cav}}_{z-1}(h)$. 

\begin{figure}
\includegraphics[width=0.6\textwidth]{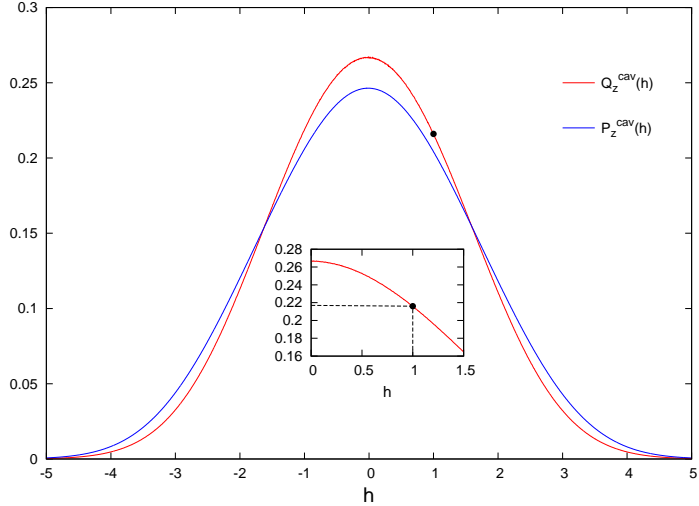}
\caption{The distribution $Q_z^{\text{cav}}(h)$ (red line) and $P_z^{\text{cav}}(h)$ (blu line) computed numerically at the zero temperature critical point on a graph with connectivity $z+1=3$. The black dot corresponds to the point $Q_z^{\text{cav}}(1) \simeq 0.217$. \label{fig:Qcavity}}
\end{figure}

The most general form of the function $P_{\ell}(J,h)$ is
\begin{equation}
P_{\ell}(J,h)=p_{\ell}Q_{\ell}(J,h)+(1-p_{\ell})\delta(J)S_{\ell}(h)\,\, \label{eq:generalP}.
\end{equation} 
For $\ell\rightarrow\infty$, the following limit must hold
\begin{equation}
\lim_{\ell\rightarrow\infty}\int \dd J\ P_{\ell}(J,h)=P^{\mathrm{cav}}_z(h)\,\, ,
\end{equation}
and then
\begin{align}
\lim_{\ell\rightarrow\infty}\ \ p_{\ell}&\ =\ 0\,\, \\
\lim_{\ell\rightarrow\infty}\ \ S_{\ell}(h) &\ =\ P^{\mathrm{cav}}_z(h)\,\, .
\end{align}
Moreover we can define the following limit
\begin{equation}
Q^{\mathrm{cav}}_z(h)\ \equiv\ \lim_{\ell\rightarrow\infty}\int \dd J\ Q_{\ell}(J,h)\ ,
\end{equation}
corresponding to the pdf of cavity fields on chains with non-zero effective coupling $J_{\ell}$.
The limiting distribution $Q^{\mathrm{cav}}_z(h)$ is different from $P^{\mathrm{cav}}_z(h)$ (see  \figurename\ref{fig:Qcavity}) and fulfils the following equation:
\begin{equation}
Q^{\mathrm{cav}}_z(h)\ =\ \overline{\delta\Big(h-h_R - \sum_{k=1}^{z-1}u_{k}-w\Big)}\,\, ,\label{eq:Qcav}
\end{equation}
where $w$ is the asymptotical cavity bias running along a path with $J_{\ell}\neq0$. The random variable $w$ is drawn from the pdf $g(w)$, which obeys the following equation:
\begin{equation}
g(w)\ =\ \frac{1}{z_g}\mathbb{E}_r\int \dd w' \ g(w')\ \delta[w-\hat{u}(w',r)]\ \Theta(1-|w'+r|)\ ,
\end{equation}
where $\hat{u}(w',r)=\text{sign}(w'+r)\min(1,|w'+r|)$, the expectation over the field $r$ is taken using the distribution $P^{\mathrm{cav}}_{z-1}(r)$, and the costant $z_g$ guarantees that $g(w)$ is properly normalized.
In the large $\ell$ limit it is reasonable to make the approximation $Q_{\ell}(J,h)\approx Q^{\mathrm{cav}}_z(h)R_{\ell}(J)$, so that the full distribution \eqref{eq:generalP} becomes:
\begin{equation}
P_{\ell}(J,h)\approx p_{\ell}Q^{\mathrm{cav}}_z(h)R_{\ell}(J)+(1-p_{\ell})\delta(J)P^{\mathrm{cav}}_z(h)\,\,\,\,\,  \mathrm{for}\,\, \ell\rightarrow\infty\,\, .\label{eq:factorP}
\end{equation}
The only unknown function is $R_{\ell}(J)$ and the distribution of the effective coupling can be exstimated as
\begin{equation}
P_{\ell}(J)=\int dh\ P_{\ell}(J,h)\approx p_{\ell}R_{\ell}(J)+(1-p_{\ell})\delta(J)\,\,\,\,\,  \mathrm{for}\,\, \ell\rightarrow\infty\,\, .
\end{equation}
Putting \eqref{eq:factorP} into Eq. \eqref{eq:recursionP} gives the recursion equation for $R_{\ell}(J)$:
\begin{equation}
R_{\ell+1}(J)=\frac{R_{\ell}(J)\int_{I} \dd h\ Q^{\mathrm{cav}}_z(h)+\int_0^1 \dd J^{\prime}R_{\ell}(J^{\prime})\int_{II}\dd h\ Q^{\mathrm{cav}}_z(h)\delta[J-(1+J^{\prime}-|h|)/2]}{\int_0^1\dd J\ R_{\ell}(J)\int_{|h|<1+J}\dd h\ Q^{\mathrm{cav}}_z(h)}\,\, .\label{eq:recursionR}
\end{equation}
Since the distribution $P_{\ell}(J)$ concentrates on the point $J=0$, a further simplification is achieved by taking into account only the lowest order in $J$ into eq \eqref{eq:recursionR} for $\ell\rightarrow\infty$. Calling
\begin{align}
\lambda&=\int_{-1}^1\dd h\ Q^{\mathrm{cav}}_z(h)\ \ ,\\
\rho&=2Q^{\mathrm{cav}}_z(1)/\lambda\ \ ,
\end{align}
eq \eqref{eq:recursionR} becomes
\begin{equation}
R_{\ell+1}(J)=\frac{R_{\ell}(J)(1-\rho J)+2\rho\int_J^1 \dd J^{\prime}R_{\ell}(J^{\prime})}{1+\rho\langle J_{\ell}\rangle_{R}}\,\, ,\label{eq:simplifiedR}
\end{equation}
where $\langle J_{\ell}\rangle_{R}=\int \dd J R_{\ell}(J) J$ satisfies the equation
\begin{equation}
\langle J_{\ell+1}\rangle_{R}=\frac{\langle J_{\ell}\rangle_{R}}{1+\rho\langle J_{\ell}\rangle_{R}}\,\, .
\end{equation}
The solution, with the initial condition $\langle J_{1}\rangle_{R}=1$, is 
\begin{equation}
\langle J_{\ell+1}\rangle_{R}=\frac{1}{1+\ell\rho}\approx\frac{1}{\ell\rho}\,\, .
\end{equation}
The recursion equation \eqref{eq:simplifiedR} for the function $R_{\ell}(J)$ reads:
\begin{equation}
R_{\ell+1}(J)=\frac{\ell}{\ell+1}\Big[R_{\ell}(J)(1-\rho J)+2\rho\int_{J}^{1}\dd J^{\prime}\ R_{\ell}(J^{\prime})\Big]\ ,
\end{equation}
whose solution is 
\begin{equation}
R_{\ell}(J)=\rho(\ell-1)(1-\rho J)^{\ell-2}[1+a^{\ell}]\,\, ,
\end{equation}
where $a=\frac{1-\rho}{1-\rho J}<1$\,\, .
The coefficients $p_{\ell}$ can be fixed via the following equation:
\begin{equation}
p_{\ell+1}=\lambda p_{\ell}[1+\rho\langle J_{\ell+1}\rangle_{R}]\longrightarrow p_{\ell}=\ell\lambda^{\ell-1}\,\, .
\end{equation}
In conclusion, the distribution of the effective coupling $P_{\ell}(J)$ is:
\begin{equation}
\boxed{P_{\ell}(J)\ =\ \rho\lambda^{\ell-1}\ell(\ell-1)(1-\rho J)^{\ell-2}+(1-\ell\lambda^{\ell-1})\delta(J)}\,\, \label{eq:PJ}.
\end{equation} 
In the large $\ell$ limit the distribution $P_{\ell}(J)$ behaves as
\begin{equation}
P_{\ell} (J)\ \approx\ \rho\ \ell^2\lambda^{\ell}\text{e}^{-\ell\rho J}+(1-\ell\lambda^{\ell})\delta(J)\ .\label{eq:P_Jexp}
\end{equation}
The previous expression shows that the relevant effective couplings different from zero are of order $J_{\ell}=O(1/\ell)$. 
The mean value of the effective coupling $J_{\ell}$ is:
\begin{equation}
\overline{J_{\ell}}=\int \dd J\  P_{\ell}(J) J=\lambda^{\ell-1}\,\, .
\end{equation}
Even if the mean value $\overline{J_{\ell}}$ is exponentially small in $\ell$, the values of $J_{\ell}$ which mostly contribute to the average are of order $1/\ell$. The asymptotical form \eqref{eq:P_Jexp} could have been predicted knowing the scaling of the moments $\overline{J_{\ell}^q}\approx\lambda^{\ell}/\ell^{q-1}$ for $\ell\rightarrow\infty$.

The knowledge of the average $\overline{J_{\ell}}$ allows us to evaluate the ferromagnetic susceptibility:
\begin{equation}
\chi_F\propto\sum_{\ell}z^{\ell}\ \overline{J_{\ell}}\sim (1-z\lambda)^{-1}\ ,
\end{equation}
which diverges when $z\lambda=1$.

The total number of pairs with effective coupling $J_{\ell}\neq 0$, for large separations $\ell\gg 1$, is
\begin{equation}
\mathcal{N}_{\ell}=\ell(z\lambda)^{\ell}\ \ \ \mathrm{for}\ \ \ \ell\gg 1\,\, .
\end{equation}

\begin{figure}
\includegraphics[width=0.6\textwidth]{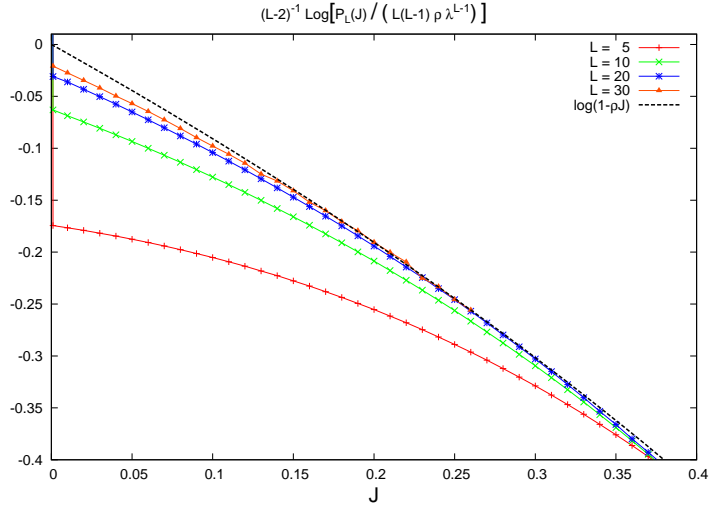}
\caption{Pdf of the effective coupling $P_{\ell}(J)$ computed numerically for $\ell = 5,10,20,30$ and the analitycal prediction given by Eq. \eqref{eq:PJ} (black dashed line). The curves are computed at the zero temperature critical point on a graph with connectivity $z+1=3$, where $\sigma_c(T=0)\approx1.037(1)$\label{fig:PJ}.}
\end{figure}

At the critical point $z\lambda=1$, so that the number of correlated not-frozen pairs (i.e. for which the response $J_{\ell}\neq 0$) is linear in the spin separation: $\mathcal{N}_{\ell}\sim\ell$.

In \figurename\ref{fig:PJ} we compare the analytical formula for the effective coupling pdf, given by Eq. \eqref{eq:PJ}, against the exact numerical estimate obtained by solving Eq.\eqref{eq:recursionP}.

In \figurename\ref{fig:Qcavity} we report the distribution $Q_z^{\text{cav}}(h)$, obtained by solving Eq.~\eqref{eq:Qcav} at the critical point. The knowledge of $Q_z^{\text{cav}}(h)$ allows us to determine the parameters $\lambda$ and $\rho$, which (at the critical point) we find to be equal to $\lambda=z^{-1}=0.5$ and $\rho\sim0.868$.

Before concluding this work let us make some observations regarding the disconnected correlation function $\overline{\langle s_0\rangle\langle s_{\ell}\rangle}$. This correlation function plays a crucial role in the analysis of the critical behaviour of the RFIM, since, in finite dimensional models, it is more singular than the connected correlation function. On a Bethe lattice the correlation functions have no singularity at all, since the correlation length never diverges. However the decay properties of $\overline{\langle s_0\rangle\langle s_{\ell}\rangle}$ at large distances are different from those of the connected correlation function. As we will see this difference is a reminiscence of the different singularities met in finite dimensional systems. 

The large distance behaviour of $\overline{\langle s_0\rangle\langle s_{\ell}\rangle}$ at zero temperature can be estimated with the following argument. Let us consider a chain in the graph connecting two spins at distance $\ell$, which we call $s_0$ and $s_\ell$. Let us consider also a third spin $s_p$ belonging to the path joining $s_0$ and $s_{\ell}$, whose distances from $s_0$ and $s_\ell$ are $p$ and $\ell-p$ respectively.

The spins $s_0$ and $s_\ell$ are correlated only if both are correlated with the spin $s_p$, that is if the effective couplings $J_{0p}$ and $J_{p\ell}$  are different from zero. We assume that both $p$ and $\ell-p$ are large enough so that $J_{0p}$ and $J_{p\ell}$ are sufficiently small. In this situation the effect of a magnetic field $H \geq \max(J_{0p},J_{p\ell})$ acting on $s_p$ will propagate both to $s_0$ and $s_{\ell}$ and thus the two boundary spins become correlated. The correlation induced by the field $H$ is of order
$\overline{J_{0p}}\;\overline{J_{p\ell}} \sim \lambda^p\lambda^{\ell-p}=\lambda^{\ell}$.
Since the position of the spin $s_p$ can be varied on the path, we have also to sum over all internal positions, thus obtaining
\begin{equation}
\overline{\langle s_0\rangle \langle s_{\ell}\rangle} \approx \sum_{p=1}^{\ell-1} \lambda^{p}\lambda^{\ell-p}\approx\ell\lambda^{\ell}\ .
\end{equation} 

The large distance decay of the disconnected correlation function $\overline{\langle s_0\rangle\langle s_{\ell}\rangle}$ contains a pre-exponential factor which grows linearly with $\ell$. In momentum space\footnote{The momentum space on the Bethe lattice can be defined via the Fourier transform $\tilde{G}(p)$ of a summable function $G(\ell)$, i.e. $\tilde{G}(p)=\sum_{\ell=0}^{\infty}G(\ell)\cos(p\ell)$.} $\overline{\langle s_0\rangle\langle s_{\ell}\rangle}$ has a Lorentzian squared behaviour $\left(p^2+\mu^2\right)^{-2}$, where the mass $\mu\propto\log(1/\lambda)$. The response function $\overline{J_{\ell}}$, instead, has a simple Lorentzian behavior $\left(p^2+\mu^2\right)^{-1}$. In our model the mass $\mu$ never vanishes, since $\lambda\leq 1/z$. However in finite dimensional models the	 mass $\mu$ vanishes at the critical point and  $\overline{\langle s_0\rangle\langle s_{\ell}\rangle}$ has a double pole at zero momentum.

\section{Summary and conclusion}
In this work we perfomed an analysis of the fluctuations of the spin-spin connected correlation function in the RFIM, using the theory of large deviations. These fluctuations are of great physical relevance, since the probability distribution of correlations is nontrivial in the thermodynamical limit. Indeed, in the limit $N\rightarrow\infty $, it becomes sharply peaked for large $\ell$, but $\mathcal{C}(\ell)$ still fluctuates wildly among samples or in a given sample among different sites. 

The major effect of the quenched disorder is that of inducing very strong correlations in a small subset of the dynamical variables. We found that precisely these rare and strong correlations dominate the ferromagnetic susceptibility. Accordingly the phase transition is mainly driven by the fluctuations of the quenched disorder, rather than the thermal noise. This is the reason why the phase transition remains of the same nature at zero temperature, where   thermal fluctuations are absent. Indeed, in the limit $T\rightarrow0$, there is still a very small number of degrees of freedom having a non-vanishing response. This is sufficient to cause the divergence of the ferromagnetic susceptibility at the critical point. In our model we found that the number of variables responding to a perturbation applied on the root spin at distance $\ell$, is not exponential (as in the finite temperature case), but only linear in the separation $\ell$.    

Another prototypical model of quenched disorder is the spin-glass Ising model. It is well known that this model, on a Bethe lattice, undergoes a phase transition at zero temperature at a finite value of the external homogeneous magnetic field. When the couplings of the model assume two possible values, say $J = \pm 1$ with equal probability, then, thanks to a gauge transformation\cite{Florent} of the coupling constants, we can easily extend the zero temperature analysis of the RFIM to the spin glass model, provided we are above the critical point. The gauge transformation sends the spin $s_i$ into $s_i\tau_i$, and the couplings $J_{ij}$ into $J_{ij}\tau_i\tau_j$, where the numbers $\tau_i$ take two possible values: $\tau_i=\pm 1$.
To determine the numbers $\tau_i$, we make the following reasoning. Let us pick up a node $i$ at random in the graph, that we call the root. In the thermodynamic limit, the neighborhood of this node will be, with high probability, a tree. We start from $\tau_i=1$ for all $i$. Now we consider all the negative couplings $J_{ij} = -1$ emanating from the root. For each of those, we perform the following transformation on the spins $s_j\to\tau_j s_j$, with $\tau_j=-1$. At this point we recompute the couplings $J$ around node $j$ and we consider the all the negative couplings $J_{jk}$ emanating from node $j$, then we set $\tau_k=-1$. We keep on transforming iteratively until we reach the leaves. Thus, the gauge transformation has the effect to propagate the frustration on the boundary of the graph. If the system has only one pure state, and so it is above its critical point, the frustration cannot generate long range correlations. As a consequence the spin on the root is insensitive to the frustration on the boundary. Said in another way, the frustration on the boundary can be substituted with uncorrelated random boundary conditions, without altering the marginal distribution of the spin on the root (or the marginals of any set of variables at a finite distance from the root). Therefore, the gauge-transformed model is equivalent to a ferromagnetic model (all $J$'s positive) in a random magnetic field. We stress that this is indeed true only above the critical point. Thanks to this equivalence all the results derived in this work immediately apply. The only modification in the formulae amounts to substitute $J$ with $|J|$ in the distribution of the effective coupling.

As a final remark we want to mention that the value of connectivity ($z+1=3$), chosen to compare the zero temperature analytical predictions, is the worst case scenario, since for larger values of $z$ the approximations used in the derivation of the effective coupling pdf $P_{\ell}(J)$ become closer to exact.

\section*{Acknowledgements}
This research has received financial support from the European Research Council (ERC) through grant agreement No. 247328 and from the Italian Research Minister through the FIRB project No. RBFR086NN1.
We also acknowledge many useful discussions with U. Ferrari, C. Lucibello, T. Rizzo and J. Rocchi.

\end{document}